\documentstyle[referee,epsf]{mn}
\input psfig.tex
\def\lsim{\lower.5ex\hbox{$\; \buildrel < \over \sim \;$}}
\def\gsim{\lower.5ex\hbox{$\; \buildrel > \over \sim \;$}}
\newif\ifAMStwofonts
\ifoldfss
  \ifCUPmtlplainloaded \else
    \NewTextAlphabet{textbfit} {cmbxti10} {}
    \NewTextAlphabet{textbfss} {cmssbx10} {}
    \NewMathAlphabet{mathbfit} {cmbxti10} {} % for math mode
    \NewMathAlphabet{mathbfss} {cmssbx10} {} %  "   "    "
  \fi
  \ifAMStwofonts
    \ifCUPmtlplainloaded \else
      \NewSymbolFont{upmath} {eurm10}
      \NewSymbolFont{AMSa} {msam10}
      \NewMathSymbol{\upi}     {0}{upmath}{19}
      \NewMathSymbol{\umu}     {0}{upmath}{16}
      \NewMathSymbol{\upartial}{0}{upmath}{40}
      \NewMathSymbol{\leqslant}{3}{AMSa}{36}
      \NewMathSymbol{\geqslant}{3}{AMSa}{3E}

    \fi
  \fi
\fi % End of OFSS
\ifnfssone
  \newmathalphabet{\mathit}
  \addtoversion{normal}{\mathit}{cmr}{m}{it}
  \addtoversion{bold}{\mathit}{cmr}{bx}{it}
  \newmathalphabet{\mathbfit} % math mode version of \textbfit{..}
  \addtoversion{normal}{\mathbfit}{cmr}{bx}{it}
  \newmathalphabet{\mathbfss} % math mode version of \textbfss{..}
  \addtoversion{normal}{\mathbfss}{cmss}{bx}{n}
  \addtoversion{bold}{\mathbfss}{cmss}{bx}{n}
  \ifAMStwofonts
    \ifCUPmtlplainloaded \else
      \UseAMStwoboldmath
      \makeatletter
      \new@mathgroup\upmath@group
      \define@mathgroup\mv@normal\upmath@group{eur}{m}{n}
      \define@mathgroup\mv@bold\upmath@group{eur}{b}{n}
      \edef\UPM{\hexnumber\upmath@group}
      \new@mathgroup\amsa@group
      \define@mathgroup\mv@normal\amsa@group{msa}{m}{n}
      \define@mathgroup\mv@bold\amsa@group{msa}{m}{n}
      \edef\AMSa{\hexnumber\amsa@group}
      \makeatother
      \mathchardef\upi="0\UPM19
      \mathchardef\umu="0\UPM16
      \mathchardef\upartial="0\UPM40
      \mathchardef\leqslant="3\AMSa36
      \mathchardef\geqslant="3\AMSa3E
    \fi
  \fi
\fi % End of NFSS release 1

\ifnfsstwo
  \DeclareMathAlphabet{\mathbfit}{OT1}{cmr}{bx}{it}
  \SetMathAlphabet\mathbfit{bold}{OT1}{cmr}{bx}{it}
  \DeclareMathAlphabet{\mathbfss}{OT1}{cmss}{bx}{n}
  \SetMathAlphabet\mathbfss{bold}{OT1}{cmss}{bx}{n}
  \ifAMStwofonts
    \ifCUPmtlplainloaded \else
      \SetSymbolFont{UPM}{bold}{U}{eur}{b}{n}
      \DeclareSymbolFont{AMSa}{U}{msa}{m}{n}
      \DeclareMathSymbol{\upi}{0}{UPM}{"19}
      \DeclareMathSymbol{\umu}{0}{UPM}{"16}
      \DeclareMathSymbol{\upartial}{0}{UPM}{"40}
      \DeclareMathSymbol{\leqslant}{3}{AMSa}{"36}
      \DeclareMathSymbol{\geqslant}{3}{AMSa}{"3E}
    \fi
  \fi
\fi % End of NFSS release 2

\ifCUPmtlplainloaded \else
  \ifAMStwofonts \else % If no AMS fonts
    \def\upi{\pi}
    \def\umu{\mu}
    \def\upartial{\partial}
  \fi
\fi

\title{SS 433: Results of a Recent Multi-wavelength Campaign} 

\author[ S.K. Chakrabarti et al.]
{\\
{\Large \bf  {\rm Sandip K.\ Chakrabarti$^{1,2}$, B.G. Anandarao$^3$, S. Pal$^{2}$, Soumen Mondal$^3$, A. Nandi$^1$,}}\\
{\Large  \bf {\rm A. Bhattacharyya$^{2}$, Samir Mandal$^2$, Ram Sagar$^4$, J. C. Pandey$^4$, A. Pati$^5$ and S.K. Saha$^5$}}\\
$^1$S.N. Bose National Center for Basic Sciences, JD-Block, Salt Lake, Kolkata, 700098, India\\
$^2$ Centre for Space Physics, Chalantika 43, Garia Station Rd., Kolkata, 700084, India\\
$^3$  Physical Research Laboratory, Navarangapura, Ahmedabad, 380009, India\\
$^4$ ARIES, Manora Peak, Nainital, 263129, India\\
$^5$ Indian Institute of Astrophysics, Bangalore 560034, India\\
e-mail: chakraba@bose.res.in, anand@prl.ernet.in,  space\_phys@vsnl.com, soumen@prl.ernet.in,
anuj@bose.res.in,\\ sagar@upso.ernet.in, jeewan@upso.ernet.in, pati@iiap.res.in, sks@iiap.res.in}

\begin{document}
\maketitle

\begin{abstract}
We conducted a multi-wavelength campaign in September-October, 2002, to observe
SS 433. We used 45 meter sized 30 dishes of Giant Meter Radio Telescope (GMRT) for radio
observation, 1.2 meter Physical Research Laboratory Infra-red telescope at Mt Abu for IR, 
1 meter Telescope at the State Observatory, Nainital for Optical photometry, 2.3 meter
optical telescope at the Vainu Bappu observatory for spectrum and Rossi X-ray Timing Explorer (RXTE)
Target of Opportunity (TOO) observation for X-ray observations. We find sharp 
variations in intensity in time-scales of a few minutes in X-rays, IR and radio wavelengths. 
Differential photometry at the IR observation clearly indicated significant 
intrinsic variations in short time scales of minutes throughout the campaign.
Combining results of these wavelengths, we find a signature of delay of about two days 
between IR and Radio. The X-ray spectrum yielded double Fe line profiles which corresponded to 
red and blue components of the relativistic jet. We also present the broadband spectrum 
averaged over the campaign duration.
\end{abstract}

%\keywords {SS 433 --- X-ray, Infra-red and radio sources --- stars: individual (SS 433) --- 
%stars: winds, outflows, mass loss }

\noindent Submitted for Publication in MNRAS.

\section{Introduction}

The enigmatic compact star SS 433 is an well studied bright emission 
line object which is known to have a companion with an orbital period 
of $13.1$d, a large disk and two highly collimated relativistic jets 
moving at $v \sim 0.26$c. Disk axis makes an angle of $\sim 78^\circ$
with the line of sight, while the jet precesses with the axis at an 
angle of $\sim 19^\circ$ (Margon, 1984) with a periodicity of about $162.15$d.  
Several observations have been carried out over the last three decades, and yet, 
the object alluded a proper identification. Most recent estimates (Hillwig et al, 2004) 
suggest that the central object could be a low mass black hole ($2.9 \pm 0.7 M_\odot$) 
with a high mass ($10.9\pm 3.1 M_\odot$) companion.  

So far, there has been a few multi-wavelength campaign in SS 433 (e.g., Neizvestnyj, 
Pustilnik \& Efrenov, 1980; Ciatti et al. 1981; Seaquist et al. 1982;  Vermeulen 
et al. 1989; Band \& Gordon 1989; and Kotani et al. 1999). However, these are
confined to two or three wavelengths. The aim of our campaign was
(a) to carry out observations in as many wavelengths as possible,
(b) to detect the nature of the short time-scale variabilities in all 
the wavelengths, (c) to obtain a broad band spectrum of this enigmatic system in order to 
model the emission processes in future. We carried out the campaign in 
Radio (1.28 GHz), in IR (J, H, and K$^\prime$ bands), in optical (B and V bands)
and in X-ray (3-30 keV) wavelengths in September-October, 2002, when the jet is more or less normal to the
line of sight and the X-ray intensity is statistically in its minimum. Given that the jet is 
produced out of matter ejected from the accretion  disk, one would expect that 
small variabilities, if present, would exist in all the wavelengths and one would hope
to correlate these variabilities in order to `follow' individual flares or knots
as they propagate through the jets.  We did observe such variabilities in 
time scales of few minutes, though, given that quite `unknown' time delays
are present between X-rays and Optical, IR or radio emitting regions, 
we found it difficult to correlate these variabilities. However, we did find a lag of two days 
between overall variation of intensities in IR and Radio wavelengths.
Our radio observation was carried out during 26th September to 6th of October, 2002.  
The IR observations were made during
25th to 29th of September, 2002. The optical photometry was made during 27th September
to 3rd October, 2002, while X-ray observation was taken only on the 27th of September, 2002. 
Optical spectra were taken on the 27th and 28th of September, 2002.
A brief report on the variabilities in radio, IR and X-rays,
observed on the 27th of September, 2002 has already been published in Chakrabarti et al. (2003). 

In the following Sections, we present the results of our multi-wavelength campaign. 
In \S 2, we briefly describe the observations and the data reduction.
In \S 3, we present our results including the lightcurve and the broadband spectrum.
Finally, in \S 4, we draw conclusions.

Our major results are as follows: (a) The short time-scale
variations are present ($2-8$ minutes) on all the days in all the wavelengths. 
We present differential photometry results in IR for all the days.
(b) The optical and X-ray spectra contain the blue and red-shifted lines which are
compatible with the kinematic model (Abell \& Margon, 1979). (c) For the first time,
we obtained the broadband spectrum over ten decades of frequency range based on contemporaneous
data. Observations  in X-rays and radio waves at regular intervals are in progress and we plan to 
present results over longer time scales in future.

\section{Observations and Data Reduction}

Table 1 gives a log of our observations during the campaign and brief remarks on 
each observation. Column 1 gives Modified Julian Day (MJD) and the date of observation, Column 2 gives the 
wave band, Column 3 gives the telescope used and its location. For the
Giant Meter Radio Telescope (GMRT), we also give the number of antennas 
working during the observation in squared brackets. Column 4 gives the duration of the
observations in seconds.

\begin{table}
\scriptsize
\centering
\caption{Observation log of SS 433}
\vskip 0.5cm
\begin{tabular}{|c|c|c|c|}
\hline
MJD (Date) & Wave &  Telescope (Location)  & Duration (s) \\
 & Band &   & \\
\hline
52542 (25/9/02) & J &   PRL(Mt. Abu)  &  1480  \\
 & H &   PRL(Mt. Abu)  & 1720  \\
 & K$^\prime$ &   PRL(Mt. Abu)  & 740  \\
\hline
52543 (26/9/02) & 1.28 GHz & GMRT(Pune) [20]$^{a)}$  &  2160  \\
&J & PRL(Mt. Abu) & 3640 \\
\hline
52544(27/9/02) & 1.28 GHz & GMRT(Pune) [28]  &  21600  \\
&J & PRL(Mt. Abu)   & 2500  \\
&H & PRL(Mt. Abu)   & 2390  \\
&K$^\prime$ & PRL(Mt. Abu)  & 2180\\
& B & State Obs.(Nainital)   & 1320 \\
& Optical & VBT (Kavalur) & 2400\\
& spectroscopy & VBT (Kavalur) & 2400\\
& 3-30 keV & RXTE   & 5696  \\
\hline
52545(28/9/02) & 1.28 GHz &  GMRT(Pune) [24]  &  960 \\
& B & State Obs. (Nainital)  & 4860 \\
& Optical & VBT (Kavalur) & 3900 \\
& spectroscopy & VBT (Kavalur) & 3900 \\
\hline
52546(29/9/02) & 1.28 GHz &  GMRT(Pune) [13] & 840 \\
& J & PRL(Mt. Abu)  & 1160 \\
& H & PRL(Mt. Abu)  & 780 \\
& K$^\prime$ & PRL(Mt. Abu)  & 475\\
\hline
52547(30/9/02) & 1.28 GHz & GMRT(Pune) [26]  & 3777  \\
\hline
52548(1/10/02) & 1.28 GHz & GMRT(Pune) [28]  & 3777  \\
\hline
52550(3/10/02) & B &  State Obs. (Nainital) & 120   \\
& V & State Obs. (Nainital) & 120 \\
\hline
52552(5/10/02) & 610 MHz &  GMRT(Pune) [29]  & 1130  \\
\hline
52553(6/10/02) & 610 MHz &  GMRT(Pune) [10]  & 2850 \\
\hline
\end{tabular}

\noindent {\bf \small $a)$} The number of antennas working during the observation.

\end{table}

Radio observations were carried out with the GMRT which has $30$ antennas each 
of $45$m in diameter in a Y-shaped array with the longest baseline  interferometry
over $25$km region (Swarup et al. 1991) near Pune, India along roughly $Y$ 
shaped array. The observations were carried out at 1.280 GHz (bandwidth $16$ MHz) 
during Sept. 26th, 2002 to Oct. 1st, 2002 and at 610 MHz (bandwidth $16$ MHz) 
during 2nd-6th October, 2002.  However, results of 3rd-4th October were full of scintillations.
The data was binned at every $16$ seconds. AIPS package was used to reduce the data. 
Bad data was flagged using tasks UVFLG and TVFLG and the standard deviation
at each time bin using UVPLT package was computed. Generally, 3C48 and
3C286 were used as the flux calibrators whenever available.

Infrared observation was made using the Physical Research Laboratory (PRL) 1.2m Mt.
Abu infrared telescope equipped with Near-Infrared Camera and Spectrograph 
(NICMOS) having 256 $\times$ 256 HgCdTe detector array cooled to the liquid 
nitrogen temperature $77$K. One pixel corresponds to $0.47$ arcsec on the sky, 
giving a field of view of $2 \times 2$ arcmin$^2$. The
filters used were standard J ($\lambda$=1.25 $\mu$m, $\Delta\lambda$= 0.30
$\mu$m), H ($\lambda$=1.65 $\mu$m, $\Delta\lambda$= 0.29 $\mu$m) and K$^\prime$
($\lambda$=2.12 $\mu$m, $\Delta\lambda$= 0.36 $\mu$m) bands.  
Short exposures were taken in immediate succession in the three bands. Single-frame
exposure time during whole observations in the J and H filters were 10
seconds. Observations in  K$^\prime$ filter were  taken with 2 sec exposure and 5 successive
frames were binned to obtain 10 sec for a better signal to noise ratio. 
On the 26th of Sept. only J band observation could be made before clouds covered the sky.
At each dithered position ten frames were taken with each integration time of $10$
seconds. The nearby infrared bright standard star GL748 (Elias et al. 1982)
were used as the flux calibrator and it was observed for 50 frames with
exposure time of 10 sec were observed in each filter during each night.

Data reduction of JHK$^\prime$ images were performed in a standard way using the DAOPHOT task
of IRAF package. All the objects and standard star frames were de-biased, sky-subtracted 
and flat fielded. The sky frames were created by usual practice of median combining of at least five
position-dithered images where the source was kept within the field of NICMOS of $2^\prime
\times 2^\prime$. At each dithered position at least 10 frames of 10 sec exposure were
taken  for J and H bands while 20 frames of 2 sec exposure were taken for K$^\prime$ band. The zero
point of the instrument was taken from the standard star observations.  
We measured the stellar magnitudes using the aperture photometry
task (APPHOT) in IRAF. Our derived mean JHK$^\prime$ magnitudes on Sept. 25th, 27th and 29th
are 9.51$\pm$0.04, 8.48$\pm$0.03 and 8.49$\pm$0.08; 9.47$\pm$0.02, 8.48$\pm$0.02 and 8.32$\pm$0.02;  
9.51$\pm$0.01, 8.49$\pm$0.04 and 8.38$\pm$0.03 respectively. On the 26th of September J magnitude
was 9.52$\pm$0.02. The magnitudes are converted
to flux density (Jansky) using the zero-magnitude flux scale of Bessell,
Castelli \& Plez (1998) for plotting purpose. 
The differential magnitudes are determined using two brightest stars 
%(std1: J= 12.1, H=10.6, std2: J=12.5, H= 11.1 mag) 
in the same frame of the object. The error in individual flux density measurement
is the usual propagation error of the observed photometric magnitude. Photometric
errors $\epsilon$ are calculated for individual frame of every star and for the subtracted
differential magnitude the final error was calculated as
$\sqrt{\epsilon_1^2 + \epsilon_2^2}$, where $\epsilon_1$ and $\epsilon_2$ are the
error-bars of the individual stars. 

The optical photometry was carried out at the State Observatory (currently known as ARIES), Nainital, India
using its 1m reflector. The photometric observations in Johnson B and V bands were carried out using 
a CCD camera at f/13 Cassegrain focus of the telescope.
The CCD system consists of $24 \times 24~ \mu^{2}$ size pixel,
having $2048 \times 2048 $ pixels. To improve the signal-to-noise ratio
the observations have been taken in a binning mode of $2 \times 2$~ pixel$^{2}$,
where each super pixel corresponds to $0.72 \times 0.72$ arcsec$^{2}$. 
The CCD covers a field of view of $\sim 13 \times 13$ arcminute$^{2}$.
Multiple CCD frames were taken with the exposure time of 120 secs. A number
of bias and twilight flat field frames were also taken
during the observing run. The frames were cleaned employing
the IRAF/MIDAS software. The magnitude of the star is
determined by using the DAOPHOT. The value of atmospheric extinction
in B pass band is $0.26$  during the observation and this was taken into account.
Due to scattered clouds, only a few exposures could be made on the 27th of September, 2002
and only one exposure on each of the B and V bands could be made on the 3rd of October, 2002.
Several exposures were taken on the 28th of September, 2002. 

The optical spectroscopic study of SS 433 was carried out with the 2.3 meter 
Vainu Bappu Telescope (VBT) at the Vainu Bappu Observatory (VBO), Kavalur,
India. CCD images were obtained during
the period of 27th - 28th September'02. Detailed description of the telescope
characteristics and observation techniques are given in Prabhu et al. (1995).
The source was pointed at for a maximum exposure time of $20$ min with the
source positioned at the center of  the CCD frame. 
The data was analysed with PC-IRAF 2.12.1-EXPORT version.
The spectrum processing comprised of several subroutines which were performed in pipe-line.
These include (a) making the MASTERBIAS using all the bias files supplied with the data
with {\emph{median combine}} option, (b) making the MASTERFLAT
using all flat files supplied with data with {\emph{median combine}} option, (c)
checking one of the flat files to get range of useful data, 
(d) using {\emph{CCDPROC}} on all the science and science
calibration files to correct for BIAS, FLAT fielding, trimming out of noise,
(e) removal of cosmic rays using {\emph{cosmicrays}} utility, 
(f) aperture synthesis of science data after
checking dispersion axis and matching with {\emph{apall}} parameters, (g) using
{\emph{apall}} for science calibration file with reference to the science data to calibrate with
the proper science data, (h) calibration of spectrum lines in calibration data, (i) wavelength
calibration (non-linear) of science data file using science calibrator file using {\emph{dispcor}}
(j) continuum calibration of wavelength calibrated science data. Since we did not have a 
standard spectrum (due to bad weather), we could not perform absolute flux calibration
and so we had to rely on the simple continuum calibrated data. Iron and Neon lines 
were used to calibrate lines. 

X-ray observation was carried out using the Proportional Counter Array (PCA) 
aboard RXTE satellite. The data reduction and analysis was performed 
using software (LHEASOFT) FTOOLS 5.1 and XSPEC 11.1. We extract 
light curves from the RXTE/PCA Science Data of GoodXenon mode.  We combine 
the two event analyzers (EAs) of 2s readout time to reduce the Good Xenon data
using the perl script {\emph make\_se}. Once {\emph make\_se} script was run on the
Good\_Xenon\_1 and Good\_Xenon\_2 pairs, the resulting file was reduced
as Event files using {\emph seextrct} script to extract light curves. Good time 
intervals were selected to exclude the occultations by the earth and South 
Atlantic Anomaly (SAA) passage and also to ensure the stable pointing. 
We also extracted energy spectra with an integration time of 16s 
from PCA Standard2 data in the energy range
$3$ - $30$ keV (out of the five PCUs only data from No. 2 and No. 3 PCUs are
added together). For each spectrum, we subtracted the background data that
are generated using PCABACKEST v4.0. PCA detector response matrices are
created using PCARSP v7.10. We perform fits to
the energy spectra in the energy range 3-27 keV with the so-called `traditional model'
for SS 433, consisting of the super-position of thermal bremsstrahlung and
Gaussian lines due to the emission from the iron atoms, modified by the
interstellar absorption.

\section{Results and Discussions}

Before we present the results of our campaign, we would like to give an overview of the 
long term behaviour of SS 433 in X-rays (Nandi et al. 2004). 
Figure 1 shows the average RXTE All Sky Monitor (ASM) data.  
The complete set of ASM data since 1996 till date have 
been averaged after folding it around $368$ days. This shows clear periodicity
which is shown in Fig. 1. This `flaring' is probably due to special alignment of SS 433
with the sun as seen by ASM. The vertical dashed
lines represent the duration of our campaign. Thus our campaign took place 
in a time-frame away from such confusing region where X-ray was generally weak.
Because of this we expected that even small variations in intensities 
would be detectable. In Chakrabarti et al. (2003) such short time variations 
have been reported. 

\begin{figure}
\vbox{
\hspace {6.0cm}
\vskip -4.0cm
\centerline{
\psfig{figure=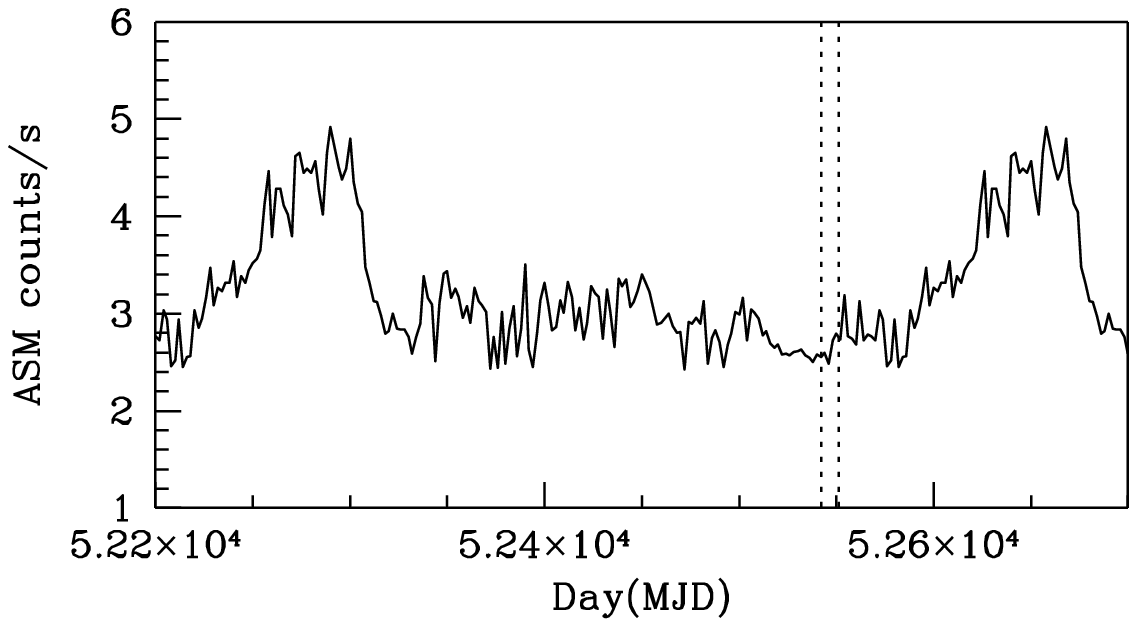,height=15truecm,width=16truecm}}}
\vskip -3.0cm
\noindent {\small {\bf Fig. 1:}
The average All Sky Monitor (ASM) light curve of SS 433 as taken by RXTE satellite. The dotted
vertical lines show the days of the present campaign, indicating that the object 
was expected to be X-ray quite. }
\end{figure}

In Fig. 2a, we present the images of SS 433 obtained by our Radio observation 
at 1.28GHz on the 1st of October, 2002. The contours are drawn at intervals of $0.055$Jy.
The beam size is shown as a circle in the lower left. In Fig. 2b, the image of 
SS 433  on the 27th of Sept. 2002, along with the those of the two standard stars in J band 
are shown. The magnitudes of the standard stars are also given.  

\begin{figure}
\vbox{
\centerline{
\psfig{figure=fig2a.ps,height=8truecm,width=10truecm}
\psfig{figure=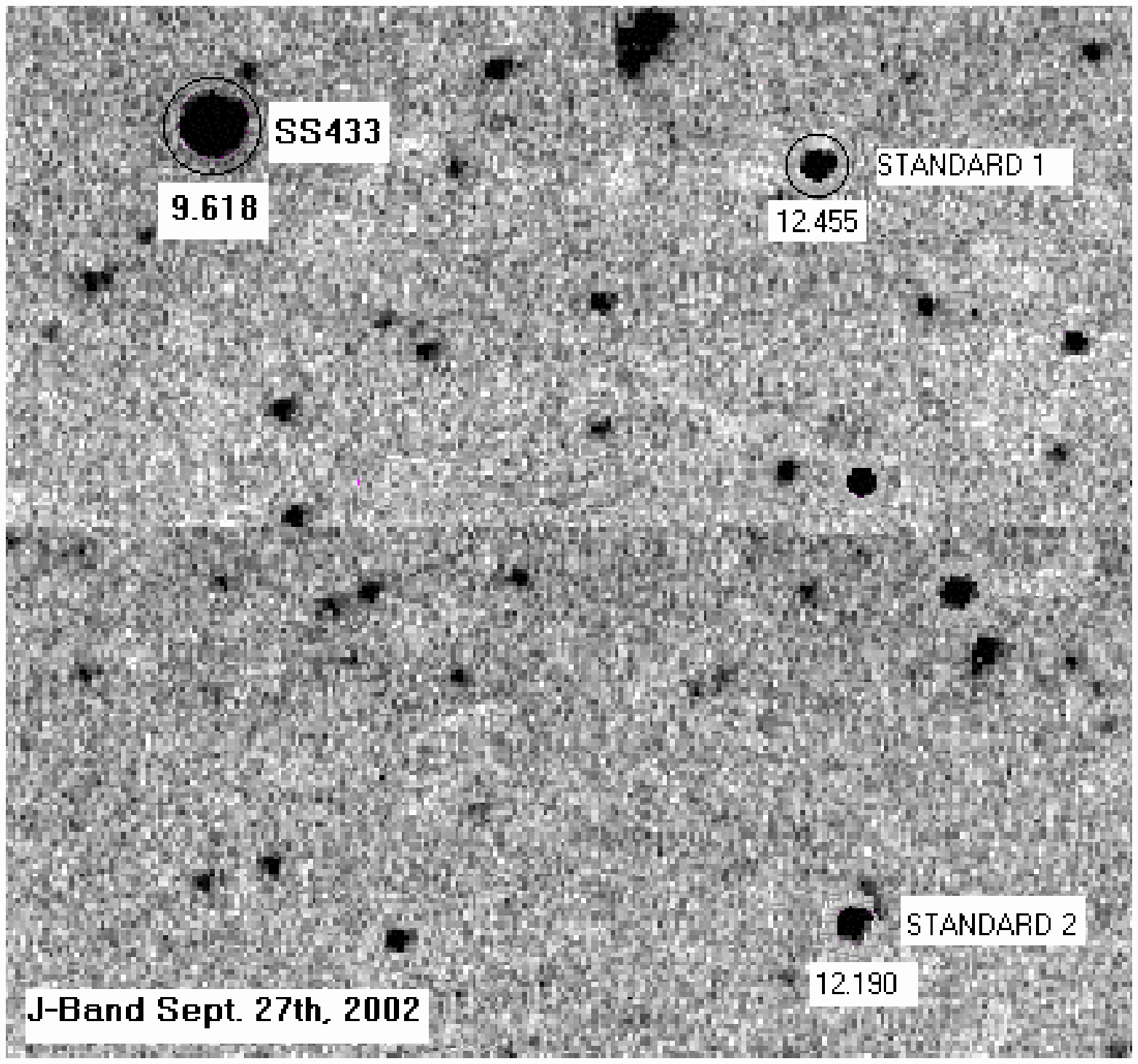,height=8truecm,width=10truecm}}}
\noindent {\small {\bf Fig. 2(a-b):}
(a) Radio and (b) IR images of SS 433. This radio image taken on 1st of October, 2002 is at 1.28 GHz. 
The peak flux is $\sim 0.55$Jy/beam. The size of the beam is given in the lower left. The
contours are at intervals of $\sim 0.055$Jy.
The IR image taken on the 27th of Sept. 2002 is in the J band. The standard stars (1 and 2)
referred in the text are also shown. The magnitudes are given as well.
}
\end{figure}

\begin{figure}
\vbox{
\vskip -4.0cm
\centerline{
\psfig{figure=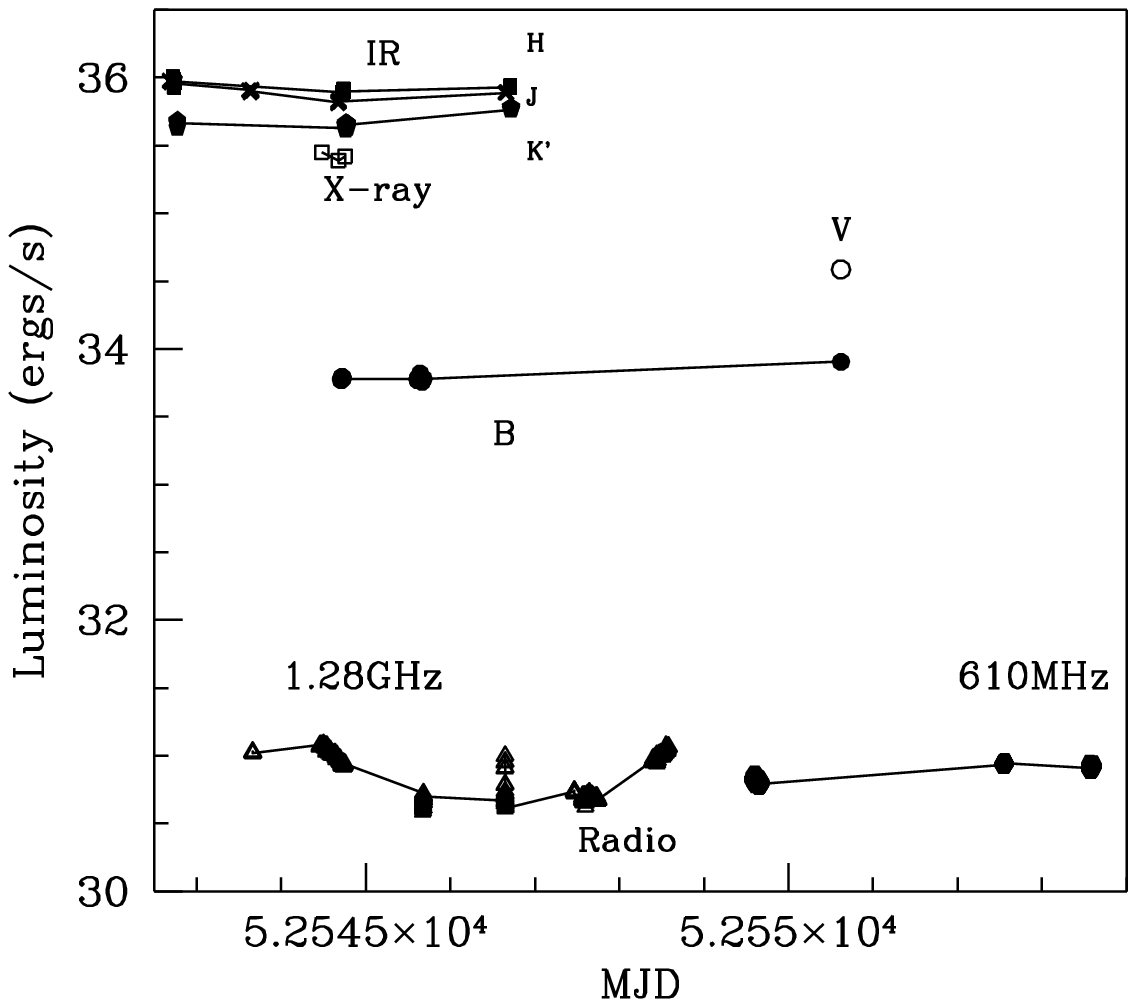,height=14truecm,width=16truecm}}}
\noindent {\small {\bf Fig. 3:}
Multi-wavelength observation of SS 433 at 1.28 GHz (triangles) band and at 610 MHz
(filled hexagons) in radio, at J (crosses), H (filled boxes), K$^\prime$ (filled pentagons)
bands in IR, B (filled circles) and V (open circle) bands in optical, and $3-25$ keV (open squares)
in X-ray during the campaign. There seems to be a lag of minimum intensity  region in radio
(MJD 52545.5 to MJD 52547.5) with respect to the Infra-red minimum region ($\sim$ MJD
52544-52545) by about 1.7 days.
}
\end{figure}

\begin{figure}
\vbox{
\centerline{
\psfig{figure=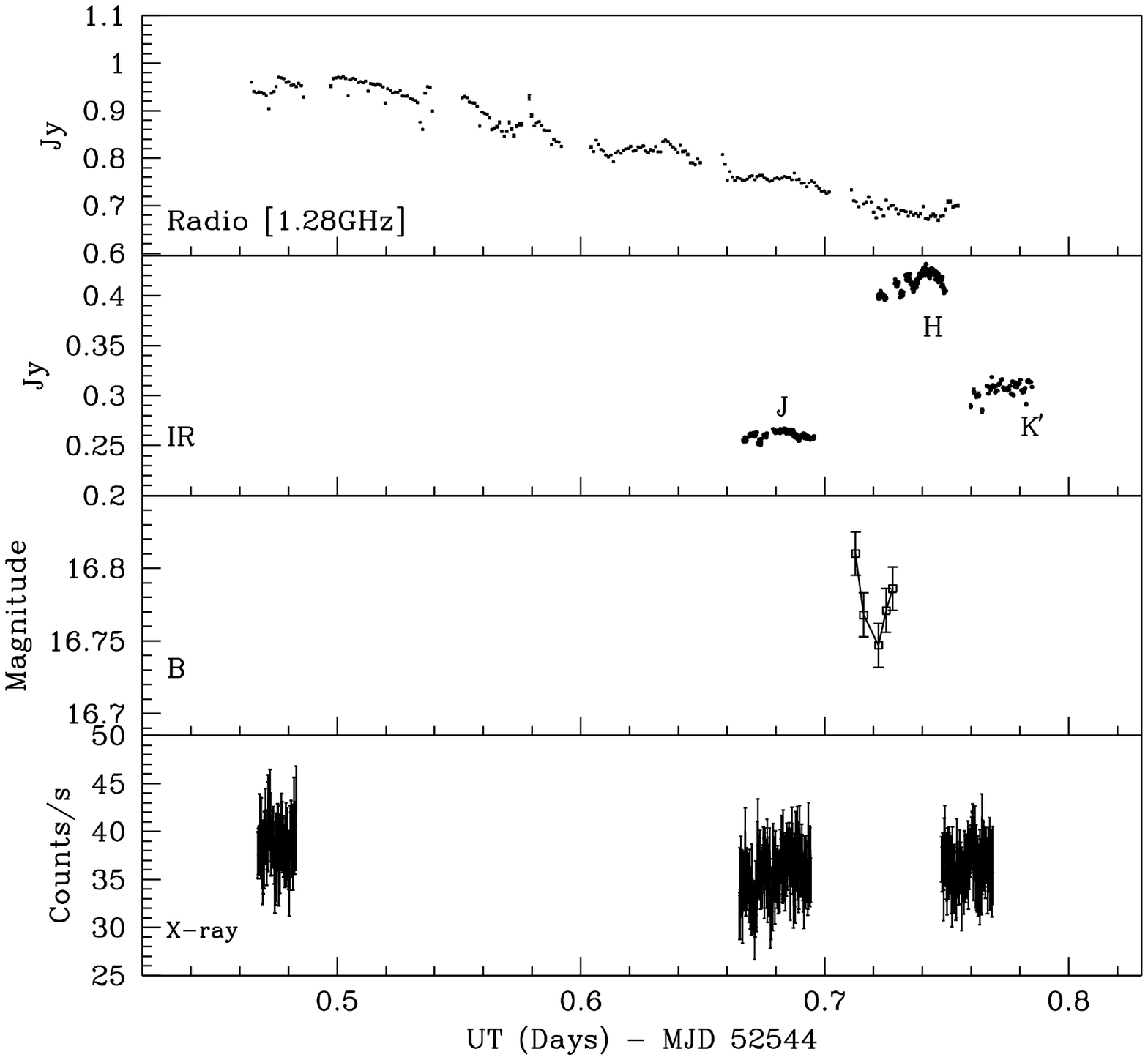,height=10truecm,width=12truecm}}}
\noindent {\small {\bf Fig. 4:}
Light curves of SS 433 on the 27th of September, 2002, as obtained by our
multi-wavelength campaign at different wavelengths.  Upper panel: Radio observation
at 1.28 GHz at GMRT, Pune. Second panel: IR observation at J, H and K$^\prime$ bands
at Mt. Abu. Third panel: B band observation at the State Observatory, Nainital
and the Bottom panel: Background subtracted X-ray count rates by RXTE satellite.
}
\end{figure}

\begin{figure}
\vbox{
\centerline{
\psfig{figure=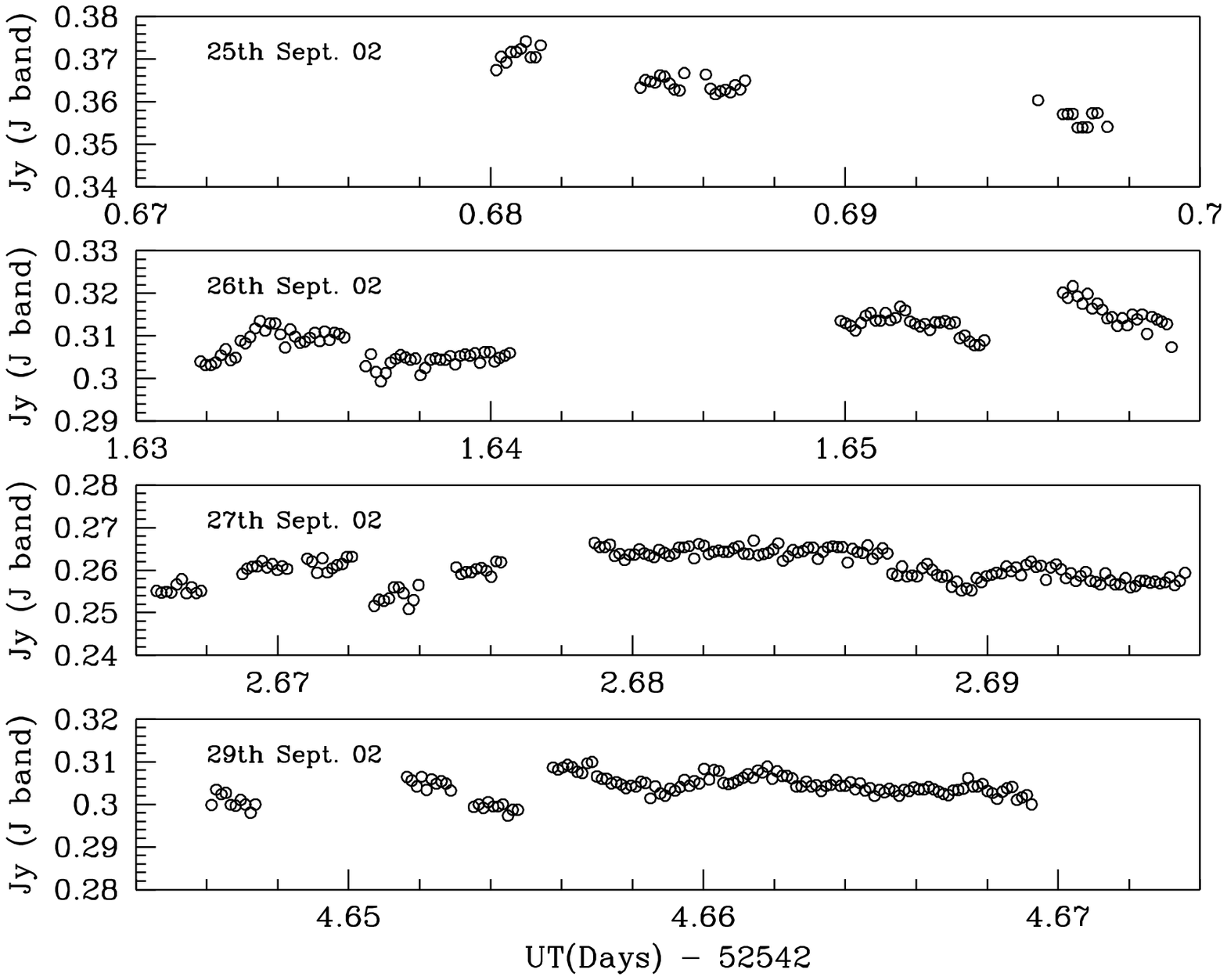,height=12truecm,width=14truecm}}}
\vskip -2.0cm
\noindent {\small {\bf Fig. 5:}
Light curves in J band as a function of the UT (Days). Considerable variations
could be seen in all the days. The size of the error-bars are well below the size of the circles.
The intensity seems to be minimum (around 0.25Jy) on  MJD 52544.674.
}
\end{figure}

\begin{figure}
\vbox{
\vskip 1.0cm
\centerline{
\psfig{figure=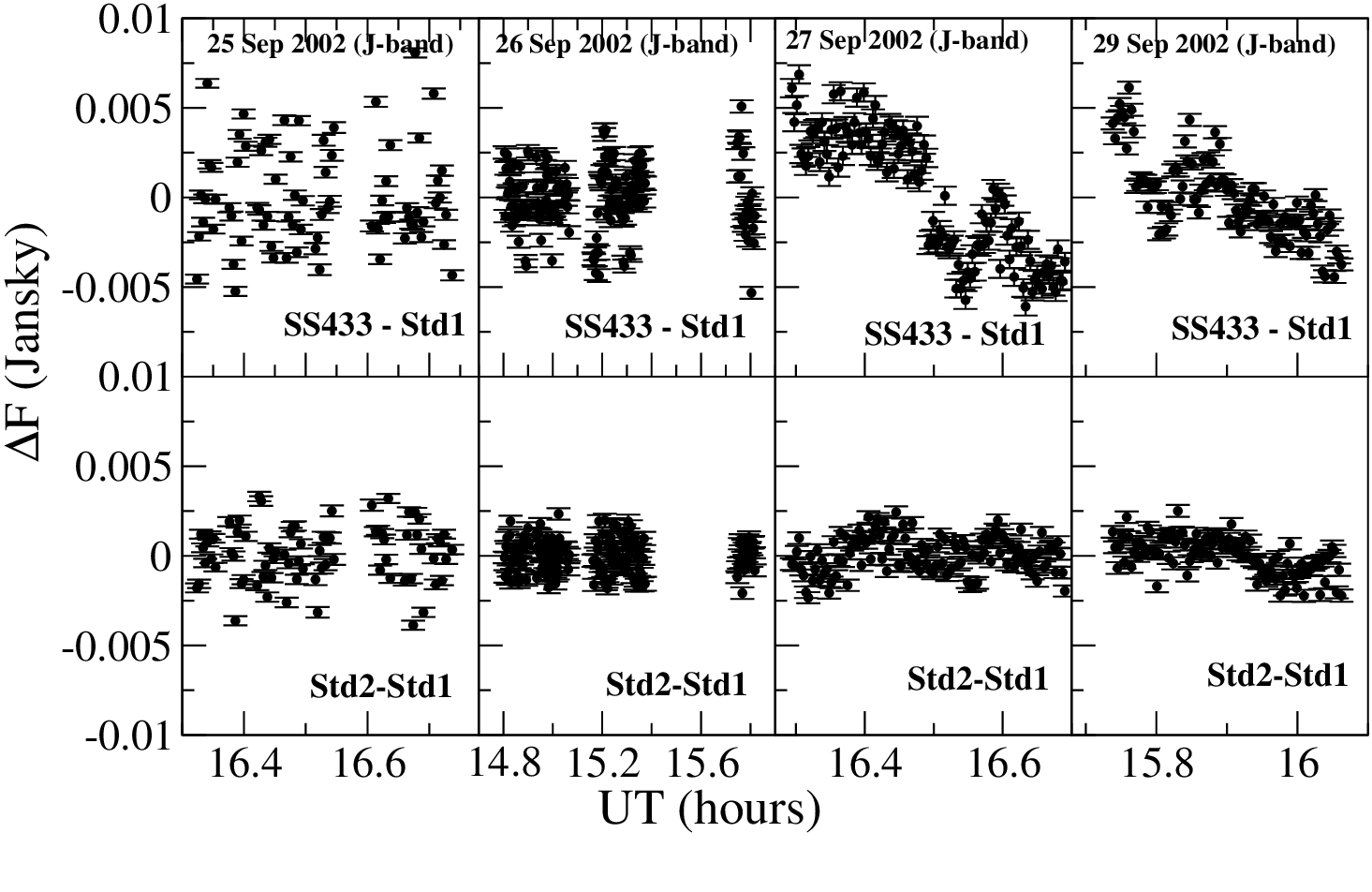,height=5truecm,width=15.5truecm}}
\vskip 1.5cm
\hskip 0.5cm
\centerline{
\psfig{figure=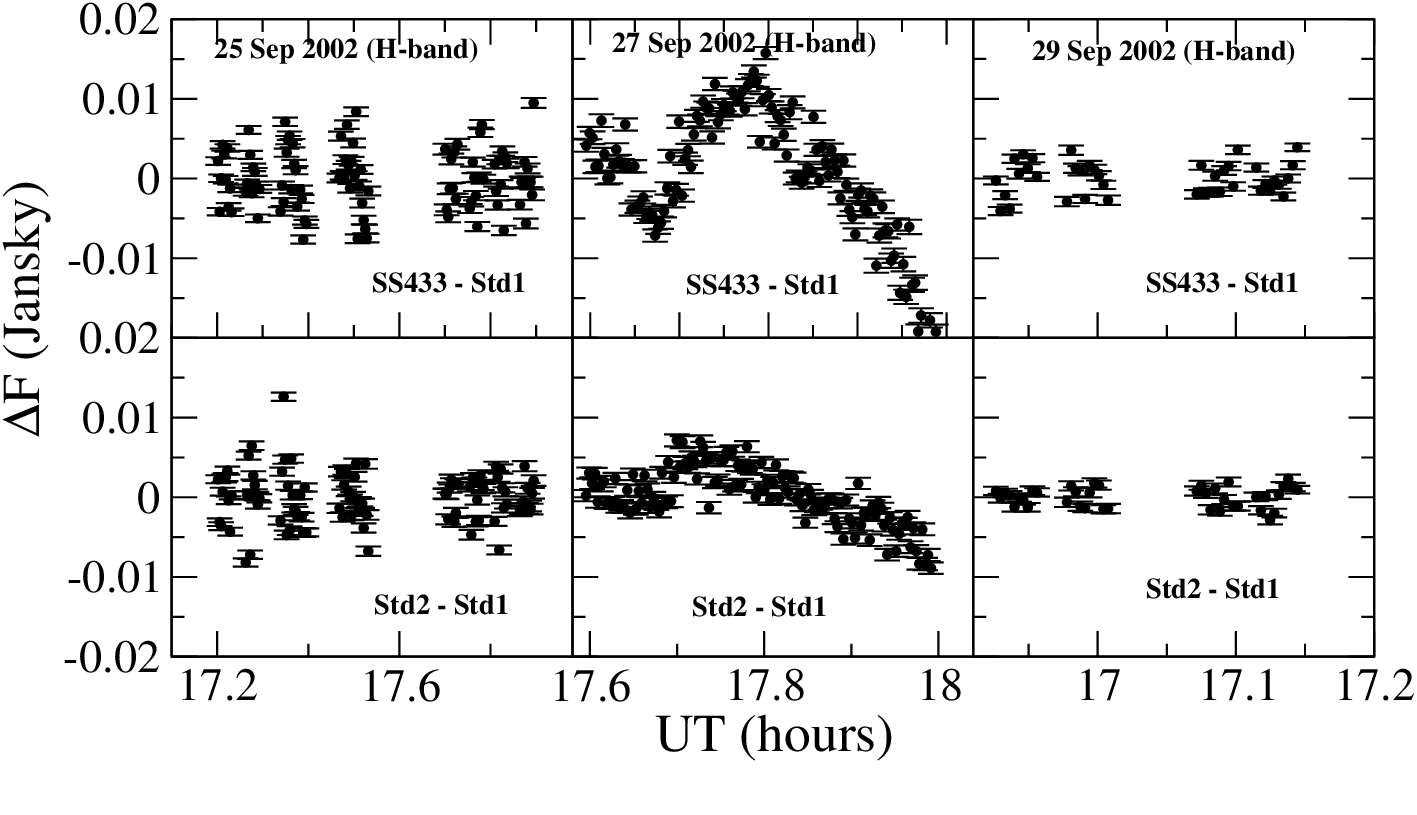,height=5truecm,width=11truecm}}
}
\vbox{
\vskip 1.0cm
\noindent {\small {\bf Fig. 6(a-b):}
Differential photometry of SS 433 in IR with respect to a standard star
(std1) in  the same frame of the object are plotted against the universal time on
various days of the campaign in the upper panels. Also shown in lower panels are the differences 
of intensities $\Delta F$(Jansky) of one standard star (std1) to the other (std2). 
The upper plots are for the data for the J band and in the lower plots are the data for the H band. 
Differential flux variation of SS 433 is above 2$\sigma$ level in comparison to 
that of the standard stars.}}
\end{figure}

Figure 3 shows the results of our multi-wavelength observation of SS 433 at 1.28 GHz 
(triangles) band and at 610 MHz (filled hexagons) in radio, at J (crosses), H (filled boxes), 
K$^\prime$ (filled pentagons) bands in IR, B (filled circles) and V (open circle) bands 
in optical, and $3-25$ keV (open squares) in X-ray during the campaign. There seems 
to be a minimum in IR data on $\sim$ MJD 52544.674 (see, Fig. 5 below) while the
radio shows minimum at $\sim$ MJD 52546.7, almost two days later. If the IR data could be taken
as the pre-cursor of the radio data, one would infer that IR was also in a state of  minimum
intensity during the campaign. However, it is to be noted that this IR intensity is the 
sum of the components coming from the companion and the jet. 
From the IR observation of Kodaira, Nakada \& Backman (1985), one notices that at the 
precessional and orbital phases of SS 433 corresponding to our campaign, the relative $K$ magnitude was 
expected to remain almost constant ($\sim  0 \pm 0.05$), while in our observation we find 
it to be highly variable ($\sim 0.225$) which suggests that there are intrinsic variation 
in IR band which may have been reflected  in the Radio band two days later. The H-band result 
was found to be higher compared to the J and K$^\prime$ band results during
the whole period. A similar result of turn-around at about $4$ micron was reported earlier 
by Fuchs (2003). This turn over could be possibly due to free-free emission in optically thin limit.
Absorption in the J band by the surrounding matter or the jets may also be a possibility.

In Fig. 4 we present the multi-wavelength lightcurves in radio (1.28GHz), in J,H, and K$^\prime$
bands, in B band and in X-rays (3-30 keV). The mean radio flux was seen to gradually go down
while behaviours in other wavebands were not so straight forward. For instance, the 
flux in $H$ band was found to be the higher compared to $J$ or $K^\prime$. This was not clearly 
understood, while a monotonic behaviour was expected. As we mentioned above, extra emission
at $H$ band is possible due to bremsstrahlung. 

In Fig. 5 we present the entire IR observations taken during the campaign. The flux was
clearly diminishing during 25th to 27th and it started rising again on the 28th.
The minimum is at around MJD 52544.674. The data is clearly variable in a few minutes
time scale. This could be seen
more clearly in the differential photometry results in the IR observations. In Figs. 6(a-b)
this is presented. In Figs. 6a (upper panel) the results in J band are presented, while in 
Fig. 6b (lower panel) the results in H band are presented. In each panel, the upper box shows the 
difference between SS 433 and the Standard 1 star while in the lower box, the difference
between the two standard stars have been plotted. The error-bars are also shown. 
In general, the upper boxes show at least twice more variation that the lower boxes. 
The 1 $\sigma$ error-bar (J=0.00035 Jy, H= 0.00085 Jy) of the differential flux
variation between SS 433 and std1 for the whole light curve is a factor of
$3.5$ and $2.5$ in the J-band and H-band respectively in comparison to that
between two standards (J=0.0001 Jy, H=0.00035 Jy). The 1 $\sigma$ for the
light curve is more than a factor of 5$\sigma$ of single point measurement
error. Thus, the variation in the IR light curves of SS 433 is likely to be intrinsic
and the analysis shows above 2$\sigma$ level variability in the both bands.
Short time-scale variation on the order of ten minutes have also been reported by Kodaira \& Lenzen (1983).

\begin{figure}
\vbox{
\centerline{
\psfig{figure=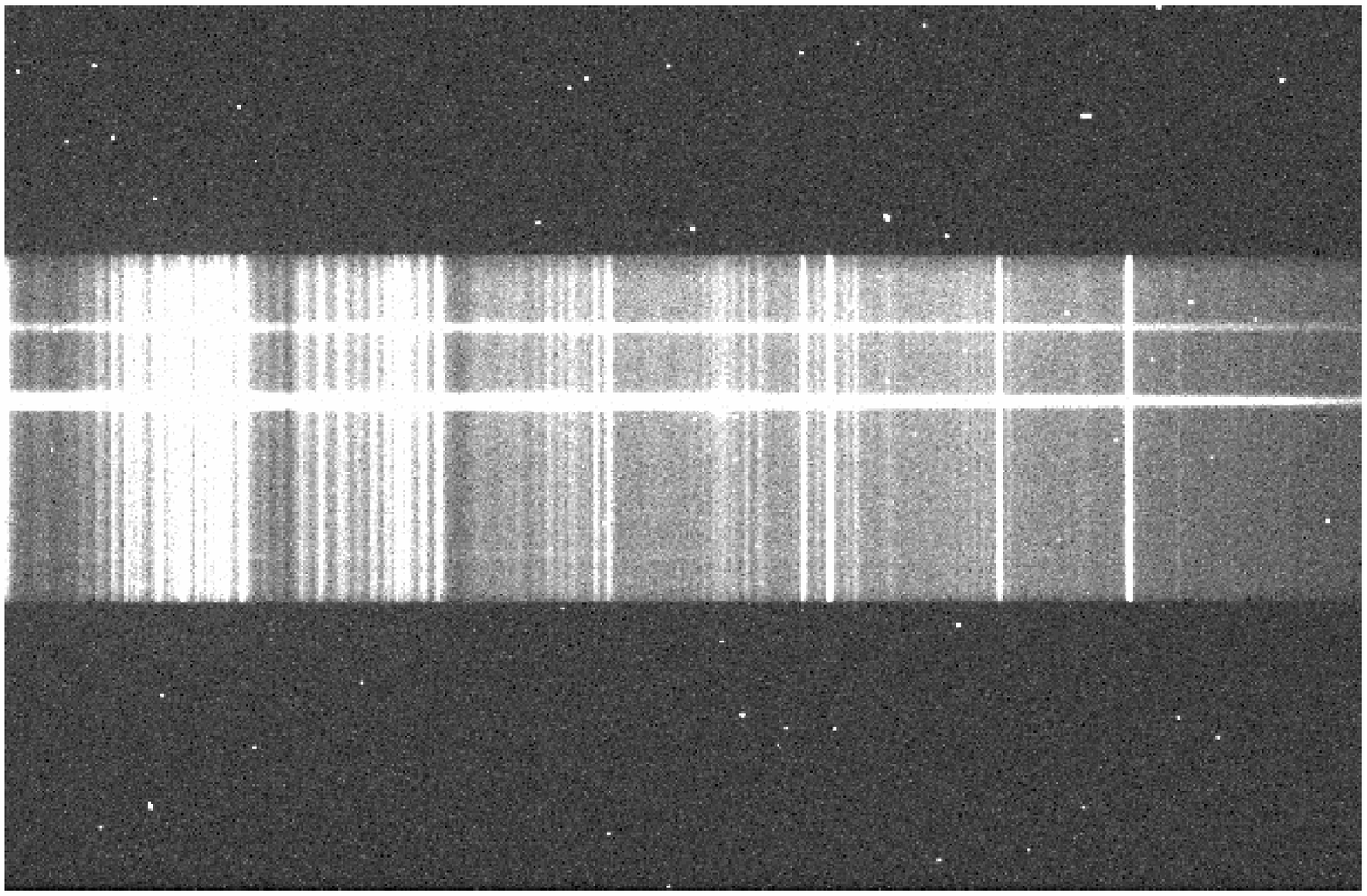,height=7truecm,width=8truecm}
\hskip 0.5cm
\psfig{figure=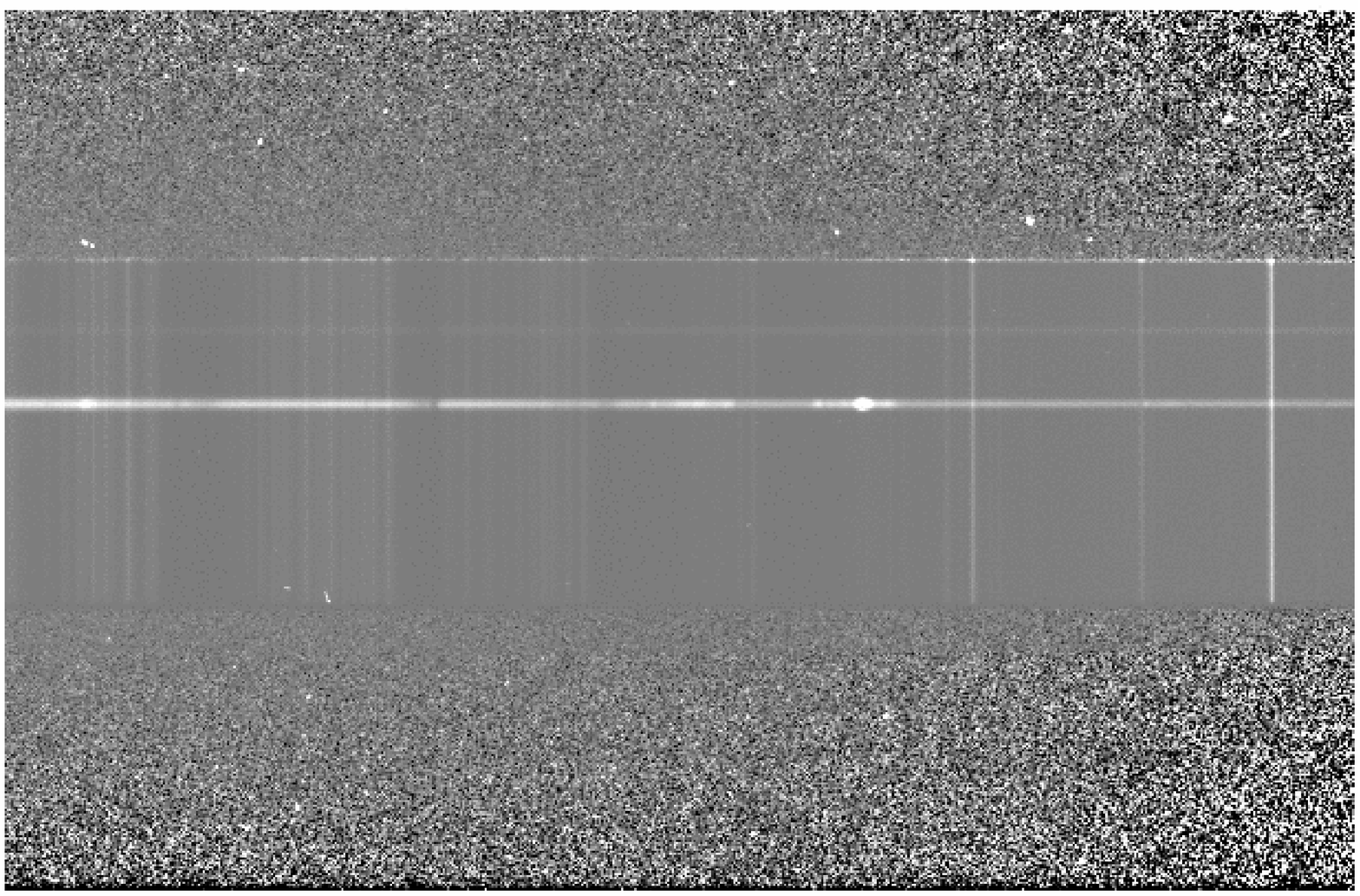,height=7truecm,width=8truecm}}}
\vskip 1.0cm
\noindent {\small {\bf Fig. 7(a-b):} a) The raw and (b) the corrected 
spectrum of SS 433 taken at the 2.3m Vainu Bappu Telescope. See text 
for detailed procedure.}
\end{figure}

\begin{figure}
\vbox{
\vskip -4.0cm
\centerline{
\psfig{figure=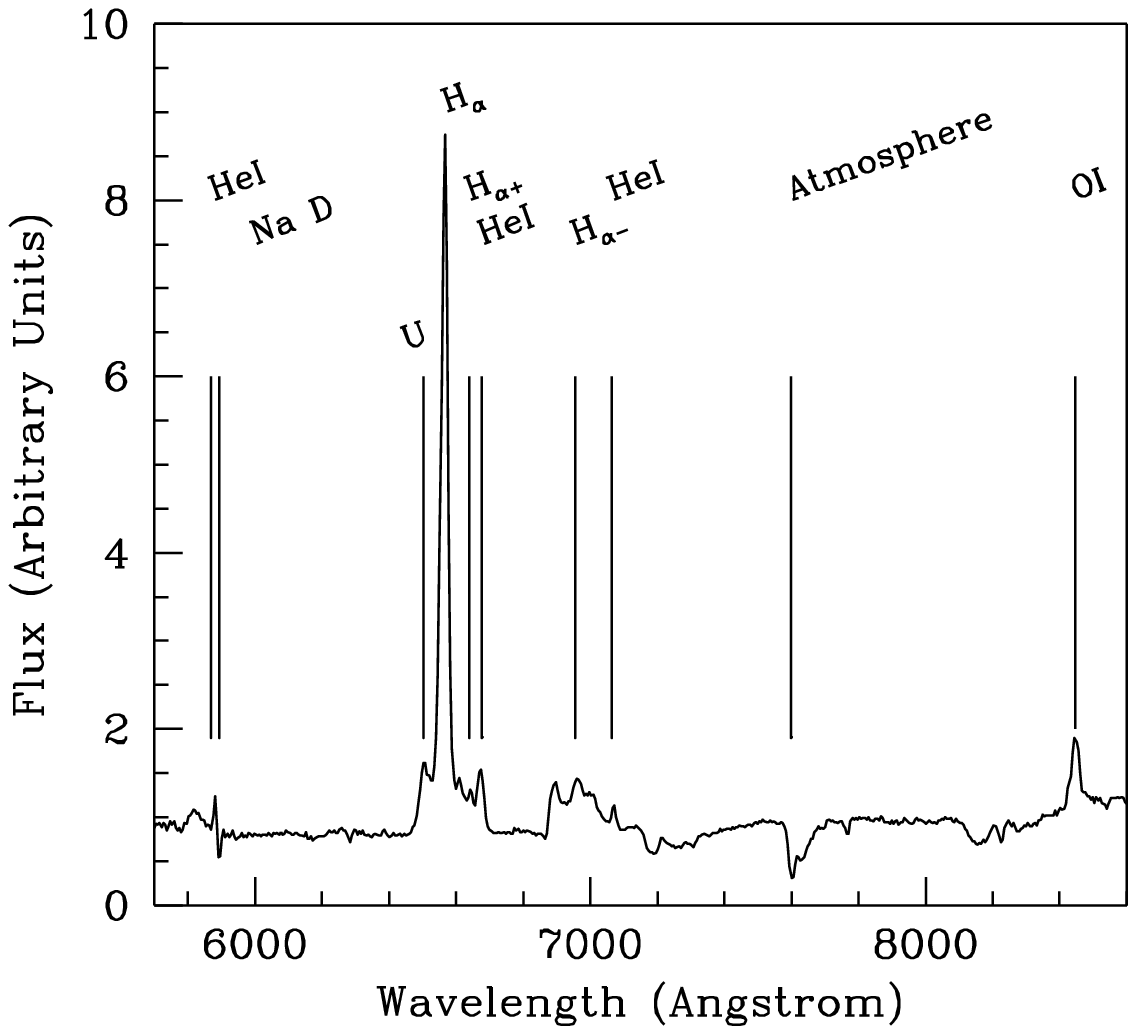,height=11truecm,width=13truecm}
\hskip -5.0cm
\psfig{figure=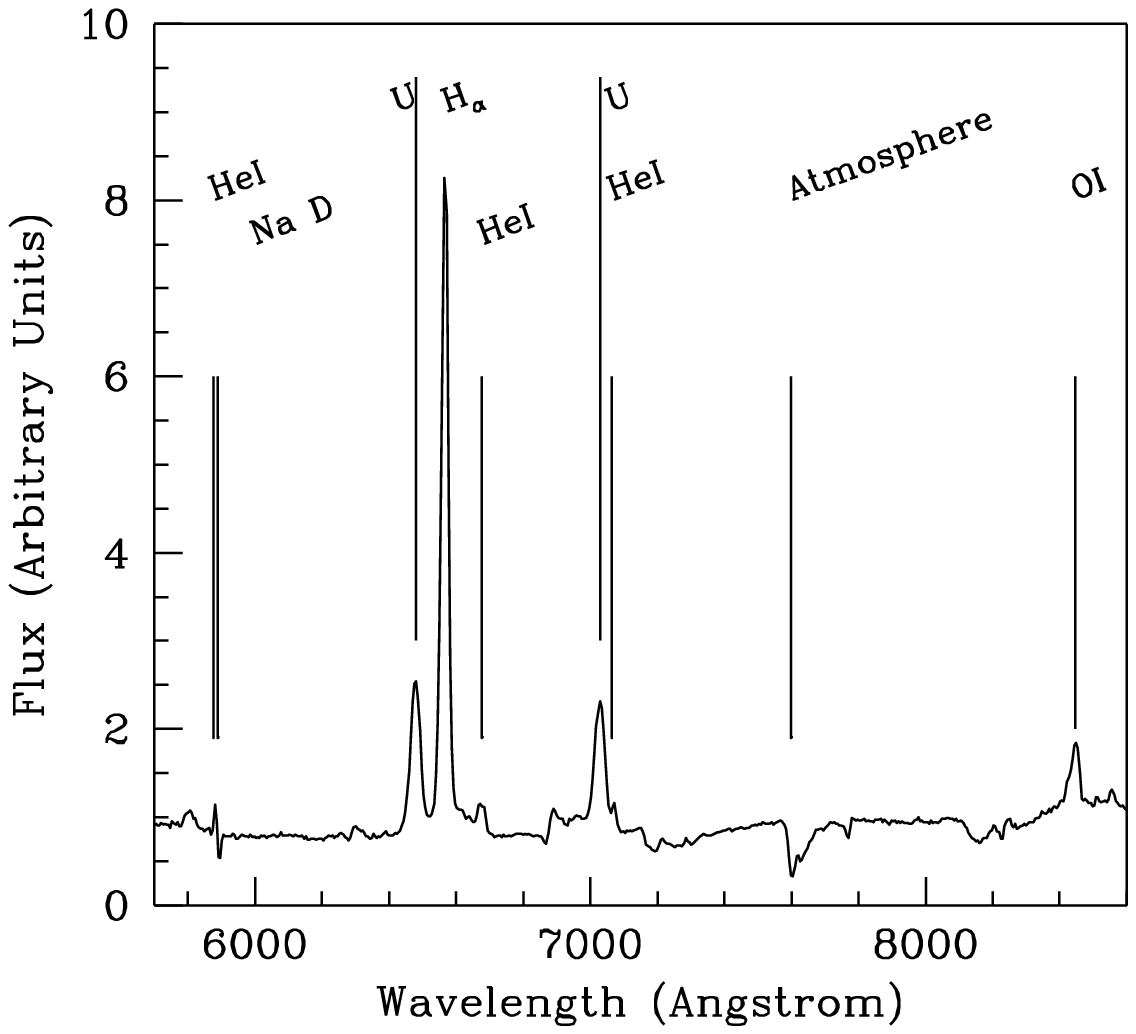,height=11truecm,width=13truecm}}}
\noindent {\small {\bf Fig. 8(a-b):} 
The calibrated, continuum subtracted, optical spectrum of SS 433 on the (a) 27th and
(b) 28th of September, 2002. In (a) the H$_\alpha$ line, blue and red shifted $H_\alpha$ lines
(denoted by H$_{\alpha+}$ and H$_{\alpha-}$ respectively), HeI lines, OI line,
atmospheric and sodium absorption lines are identified. The Doppler shifted $H_\alpha$ lines are exactly where
there are expected from kinematic model of Abell and Margon (1979) within the instrumental
resolution of $5$ \AA. In (b) we also see two bright lines, marked by `U' at $7029.07$\AA $\ $ and $6481.09$\AA $\ $ 
respectively.}
\end{figure}

\begin{figure}
\vbox{
\centerline{
\psfig{figure=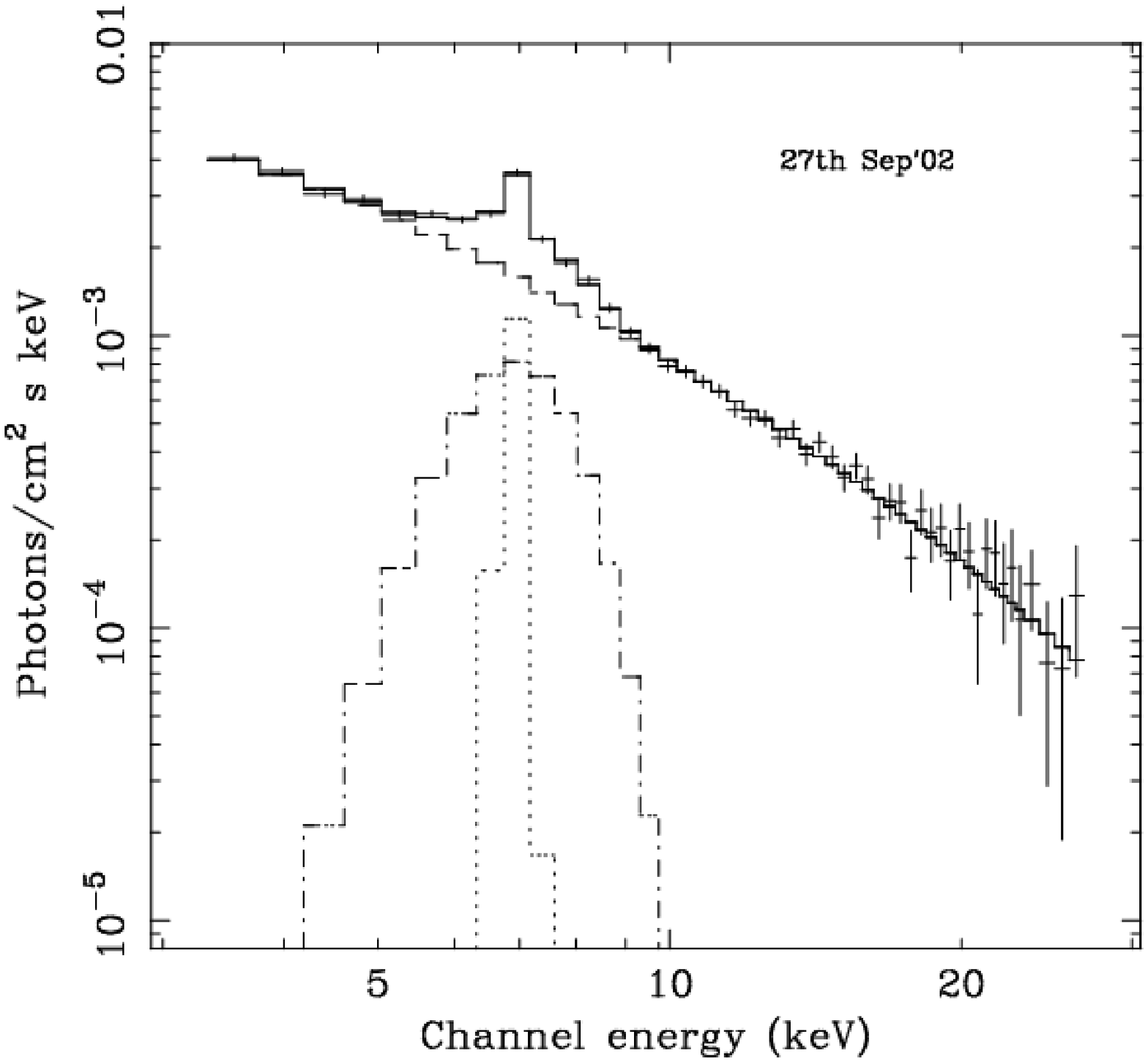,height=8truecm,width=10truecm,angle=0}}}
\vskip 1.0cm
\noindent {\small {\bf Fig. 9:} 
The X-ray spectrum of the first spell of the RXTE observation of the 27th of Sept., 2002.
The spectrum was fitted with a bremsstrahlung and two iron lines showed with dotted  and dot dashed
curves.
}
\end{figure}

\begin{figure}
\vbox{
\vskip -3.0cm
\centerline{
\psfig{figure=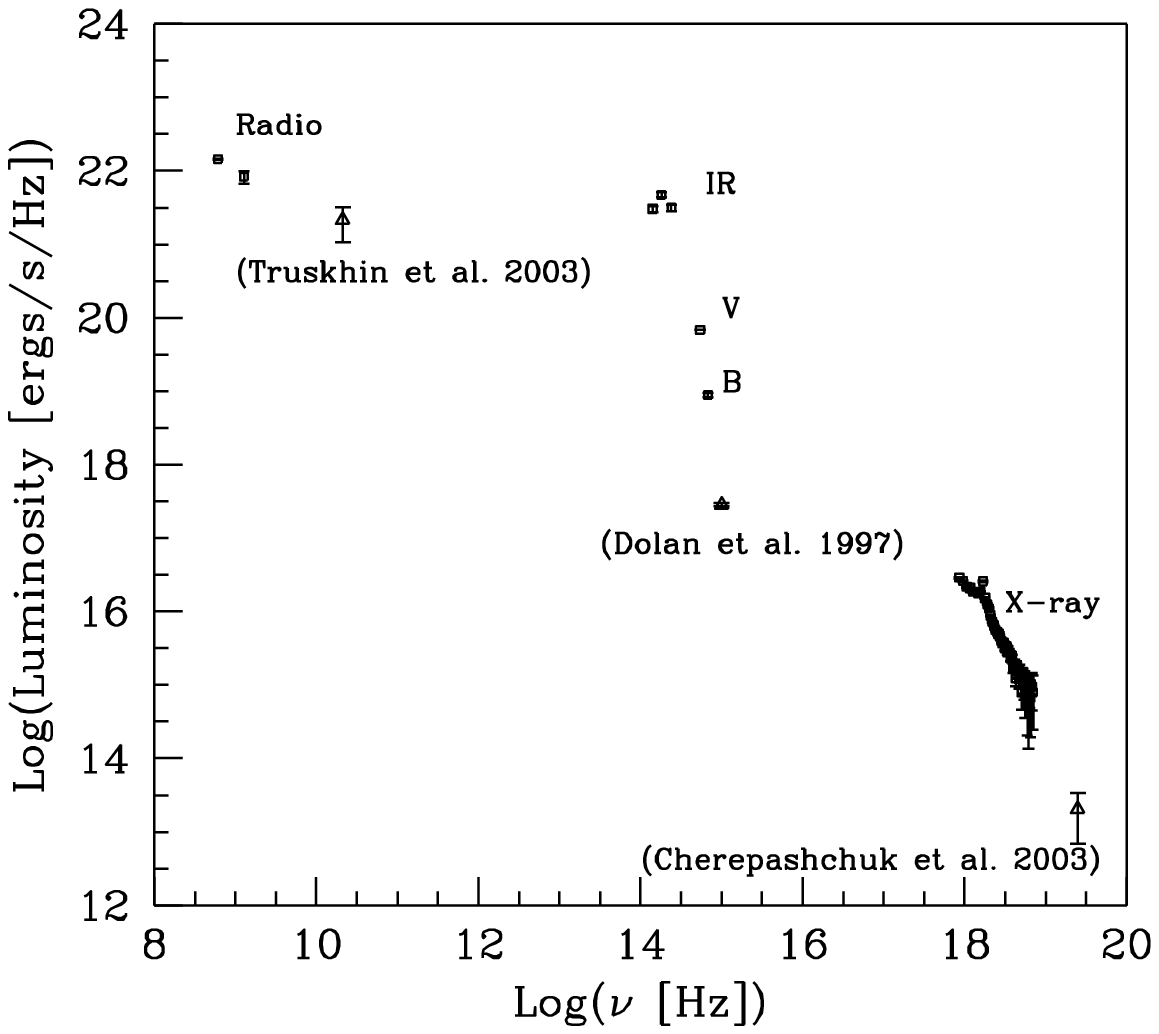,height=12truecm,width=15truecm}}}
\noindent {\small {\bf Fig. 10:} 
The multi-wavelength spectrum of SS 433 as obtained by our campaign. Here, average luminosity
(open boxes) over our available data has been plotted and the wavebands are marked. For comparison,
we included three points, marked by open triangles with error bars, from literature (marked) which are
not contemporaneous with our observation.
}
\end{figure}

Since the optical results depended heavily on the local sky conditions, the data acquisition
was not uninterrupted. Indeed, although two observatories, one in southern part of India (VBT, Kavalur)
and the other in the northern part of India (ARIES, Nainital) were chosen, both observations
were affected by the late monsoon activities. Table 1 showed the duration of the data 
acquisition and the exposure time taken. In Figs. 7(a-b) we present the spectrum of SS 433
taken at VBT. Fig. 7a shows the raw spectrum, while Fig. 7b shows the spectrum on the 27th of Sept. 2002
corrected as described in Sec. 2. In Fig. 8a, the wavelength calibrated spectrum
is shown and major lines were identified. On this date, both the jets were showing 
redshifts. The shifted line wavelengths match with what is expected from the  kinematic model of
Abell and Margon (1979) within the instrument resolution of $5$\AA. For instance,
at 13:49 UT, on 27th of Sept. 2002, the expected red shifts were $0.05901$ and $0.01234$
respectively and the $H_{\alpha-}$ and $H_{\alpha+}$ lines were expected 
at $6950$\AA $\ $ and $6644$\AA $\ $  respectively. Our observed lines were at 
$6961$ \AA $\ $ and $6642$ \AA $\ $ respectively. The spectra on the 28th
 of September, 2002, presented in Fig. 8b, had two bright lines 
(marked by `U' on the figure) at $7029$\AA $\ $ and $6481$\AA $\ $ respectively apart from 
the usual lines. The red/blue shifted $H_\alpha$ lines had very low intensity, 
indicating the decaying phase of the so-called optical bullets (Margon 1984; Vermeulen et al. 1993). 
On the 28th, the spectrum was taken 6 times, and in all the spectra these two unidentified lines
were seen. The origins of the brighter `U'-marked lines are not totally clear as they are not close 
the expected $H_\alpha\pm$ lines expected on that day. On the 27th of September, 2002,
there was an un-identified line on the blue-ward side of the $H_\alpha$ line at $6503$\AA
 $\ $as well. This is also marked as `U' in Fig. 8a. These could be 
from the accretion streams or from the winds from of the companion itself. If correct, and are identified as
the blue and red-shifted lines of the $H_\alpha$ line from the companion, then, the projected velocity
components long the line of sight  required to produce these lines would be  $21,300$ km s$^{-1}$ 
(away from the observer) and $3748$ km $s^{-1}$ (towards the observer) respectively. 
Abnormal activities in the jets may not be ruled out either, in which case the 
asymmetry of the red and blue-shifts would be due to probable intrinsic 
red-shifts of the relativistic system. Unfortunately the spectrum on the 30th of September, 2002
could not be used, as there was focusing error in the telescope.

The three spells of X-ray data (see, Fig. 4, bottom panel)  were analyzed separately. 
We found that generally speaking, two line model with a thermal bremsstrahlung is 
necessary for statistically and physically acceptable fit to the X-ray spectra. 
However, the significance depends on the duration of observation. In Fig. 9, we present the X-ray spectrum 
of the first spell of our RXTE observation. The single line fit to the spectra yields $\chi^2/\nu =
44.94/45$ ($\nu$ is degrees of freedom) whereas the double line fit gives more
acceptable value with $\chi^2/\nu = 34.84/43$. The requirement of the 2nd
line in the spectrum is tested from the {\it ftest task} within XSPEC and found
that the fit is significant at (2.4$\sigma$) level. 
%(ftest 34.83793  43 44.94129 45 F statistic value =   6.24 and probability   =  4.190E-03)
The best fitted two lines found at energies 6.966 keV and 6.898 keV
correspond to the Doppler-shifted values of z= -0.00014(FeXXVI) and
z= -0.032 (FeXXV) or +0.0096(FeXXVI) respectively. Therefore, both the lines
that we have identified are coming from the blue-jet of SS 433. The observed
flux is $2.6524 \times 10^{-10}$ ergs cm$^{-2}$ sec$^{-1}$ in the energy range 3-25 keV with
the bremsstrahlung temperature at $15.77$ keV. We failed to fit with a 
model having a blackbody emission component. Thus, no evidence for a Keplerian disk was found.
Similarly, the single line fit to the spectra of the second spell yields a $\chi^2/\nu =
54/45$ whereas a double line fit gives more acceptable value with $\chi^2/\nu = 35.7/43$. 
The requirement of the 2nd line in the spectrum is tested and found
that the fit is significant at ( 3$\sigma$) level. The best fitted two lines found 
at energies 7.012 keV and 6.802 keV which correspond to the Doppler-shifted values of z= -0.007(FeXXVI) 
and z= -0.018 (FeXXV) respectively. Here too, both the lines are from the blue-jet of SS 433. 
The observed flux is $2.375 \times 10^{-10}$ ergs cm$^{-2}$ sec$^{-1}$ in the energy range $3-25$ 
keV with the bremsstrahlung temperature at $13.92$ keV.

Figure 10 gives the broadband spectrum of SS 433 that we obtained using our multi-wavelength
campaign. Campaign average data has been used for simplicity. In radio, results of 610 MHz and 1.28 GHz
data have been put, while in infra-red, the results of
I,J, $K^\prime$ bands have been put. We included $V$ and $B$ band observations 
which clearly show heavy extinction and the luminosity drops dramatically. X-ray spectrum in 
$3-27$ keV is also shown. To compare with the results of others, we have included three observations 
at wavelengths which were not covered during our campaign (triangles). 
Thus, $21.7$GHz observation of Trushkin et al (2003), ultra-violet observation of 
Dolan et al. (1997) and gamma-ray observation of Cherepashchuk et al.
(2003) have been included. These three points were not contemporaneous to our campaign, but yet,
they generally fall at reasonable values in the overall spectrum.

\section{Concluding remarks}

In this paper, we have presented results of a recent multi-wavelength campaign on SS 433 carried out
during 25th September, 2002 till 6th of October, 2002 using radio, IR, optical and X-ray  instruments.
Average ASM result indicated that the campaign was conducted when X-ray intensity was generally low which we 
also verify from our observations. 
We found that there is a tendency for the radio intensity  variation to lag the IR variation
by about two days. The broadband spectrum clearly showed evidence of very high extinction
in optical region, possibly due to large scale obscuration of the central object by matter
coming from the companion wind (Paragi et al. 1999). 
The X-ray data could be fitted with two Fe lines, both of which
appear to be coming from the blue-shifted jet, i.e., the jet pointing towards us. 
We also find very small time-scale (few minutes) variations in all the wavelengths which could be
suggestive of small bullets propagating from the base of the jet on the accretion disk
to the radio regions as originally suggested in earlier communications (Chakrabarti 2002, 2003). 
The differential photometry in IR gives very clear indications that these short time scale
variations were intrinsic to the jet. In optical spectra we observed that the blue- and red-shifted 
$H_\alpha$ lines to appear at the same location on the 27th of Sept. 2002 -- on the
28th the intensities of these lines  were very low possibly because the so-called optical bullet emission
was on the decaying phase.  We also see two bright, unidentified
lines in all the frames taken on the 28th Sept. which could be due to sudden winds in the companion. 
Our optical and X-ray spectra indicated that the kinematic model generally gives 
the correct description even today. 

SS 433 has always been a puzzle and it is still so even after 25 years of its discovery. 
The mass estimate of the recent observation (Hillwig et al. 2004) is an indication 
that the central compact object could be a small mass black hole. Given that in the last decade
fresh understanding about the accretion processes onto black holes have emerged, 
it may be necessary to look into this system with a fresh look. Our result gives some idea about the
physical processes that is going on near the compact object (a) The X-ray spectrum does not show
any indication of a Keplerian disk. Thus the flow may be totally sub-Keplerian or advective
(Chakrabarti, 1990), as is expected if the accretion is from winds. (b) Broadband spectrum
indicated heavy extinction in the optical/UV region. This is expected from the
matter that is accumulated around the system (Paragi et al, 1999). (c) Short-time scale variations
may indicate ejection of bullet-like entities, which could be formed due to shock oscillations in the 
advective flows (Chakrabarti et al. 2002). (d) There is a general indication that the radio intensity
follows the IR by about two days. This could not be rigorously confirmed as the coverages in the campaign
were not very high. If correct, and if we assume that the same matter generally propagates from the
IR jet to radio jet, this would indicate that the radio emission takes place 
only about $6.73 \times 10^{14}$ cm away from IR emitters. This seems to be quite reasonable. In fact,
while comparing the results of Kodaira, Nakada \& Backman (1985) with our observation,
we conclude that there are intrinsic variation in IR band which may have been reflected in the Radio band
two days later.

According to Brinkmann et al. (1991), the length of the  X-ray jet is smaller than $\sim 10^{11}$ cm.
Marshall et al (2002), further quantified  this limit at $2 \times 10^{10-11}$ cm. In our analysis,
we find the temperature of X-ray emitting region to be 
$k T \sim 15$ keV   which, using the model of Kotani et al. (1996), correspond to even smaller distance
of $\sim 10^{10}$ cm. While analysing RXTE data of SS 433 for over two years, Nandi et al. (2004) 
found the temperature to be even higher.  Therefore, the X-ray emission must be from a region
much closer to the compact object. On the other hand, IR emission of the jet may be emitted
some where around $10^{12-13}$ cm (Fuchs 2003). Thus, the time lag  between X-rays and IR should 
be at most a few hundreds of seconds. Lack of continuous X-ray observation during our campaign
does not allow us to make definite comment on whether this lag was observabed or not.

SKC thanks Dr. J. Swank of NASA/GSFC for  giving us TOO time in RXTE.
We thank the staff of the GMRT who have made these observations possible. GMRT is run by the
National Centre for Radio Astrophysics of the Tata Institute of Fundamental Research.
This work is supported in part by CSIR fellowship (SP) and a DST project (SKC and AN). 
and the team of GMRT/NCRA for giving

{}

\end{document}